\documentclass{iopjournal}

\usepackage{amsmath}
\usepackage{amssymb}
\usepackage{graphicx}
\usepackage{subfig}
\usepackage{dcolumn}
\usepackage{comment}
\usepackage{soul}
\usepackage{bm}
\usepackage{xcolor}
\usepackage{natbib}
\usepackage[utf8]{inputenc}
\usepackage[T1]{fontenc}
\usepackage{mathptmx}
\usepackage{ragged2e}

\begin{document}

\articletype{Article type} 

\title{Nitrogen-doped WRe Superconducting Nanowire Single Photon Detectors}

\author{F.~Colangelo$^{1,\,2,\,3}$\orcid{0009-0005-9434-7607}, Abhishek~Kumar$^{1,\,2}$\orcid{0000-0002-7462-7442}, H.~Wang$^{3}$\orcid{0009-0008-8487-7198}, F.~Avitabile$^1$\orcid{0009-0009-3926-9609}, C.~Attanasio$^{1,\,2}$\orcid{0000-0002-3848-9169}, I.~Esmaeil~Zadeh$^3$\orcid{0000-0002-3833-2508} and C.~Cirillo$^{2,\,*}$\orcid{0000-0001-8755-4484}}

\affil{$^1$Dipartimento di Fisica ``E.R. Caianiello'', Università degli Studi di Salerno, I-84084 Fisciano (Sa), Italy}

\affil{$^2$CNR-SPIN, c/o Università degli Studi di Salerno, I-84084 Fisciano (Sa), Italy}

\affil{$^3$Department of Imaging Physics (ImPhys), Faculty of Applied Sciences, Delft University of Technology, Delft 2628 CJ, The Netherlands}

\affil{$^*$Author to whom any correspondence should be addressed.}

\email{carla.cirillo@spin.cnr.it}


\justifying

\begin{abstract}
Nitrogen-doped Tungsten--Rhenium superconducting alloys were recently proposed as a promising material platform for superconducting nanowire single-photon detectors (SNSPDs), offering a favorable balance between high normal-state resistivity and tunable superconducting properties. In this work, we report on the fabrication and characterization of SNSPDs based on thin WRe films deposited by DC magnetron sputtering in a mixed Ar/N$_2$ atmosphere. Meander detectors with 70~nm linewidth exhibit saturated internal detection efficiency (IDE) up to $1310$~nm {and $85.3\%$ IDE at $1550$~nm} at $2.5$~K, with sub-nanosecond rise times, decay times of the order of a few nanoseconds, and timing jitter of $72.3$~ps measured with room temperature amplifiers.
\end{abstract}

\section{Introduction}

Superconducting nanowire single-photon detectors (SNSPDs)~\cite{Iman2021} have become a central technology for quantum communication, sensing, and metrology, while increasingly enabling applications ranging from atmospheric monitoring to biomedical diagnostics~\cite{You,Hao2024,Salvoni2022,Ozana,Tamami}. As these fields expand, the need for detectors combining high system detection efficiency, low timing jitter, and operation at elevated temperatures continues to stimulate intense materials research. Early SNSPD implementations relied primarily on polycrystalline nitrides such as NbN and NbTiN, whose relatively high superconducting transition temperature ($T_{\rm c}$) and robustness supported rapid technological adoption~\cite{Iman2021,Wang2017,Zichi2019,Cheng2020,Shan2021,Chang2021_APLPhotonics,Chang2022,Stepanov2024}. The subsequent introduction of amorphous superconductors, including WSi and MoSi, showed that increased disorder can substantially improve the internal saturation efficiency (IDE), although often at the expense of slower response and lower operating temperatures~\cite{Korneeva2014,Zhang2016}. The availability of different SNSPD materials allows for tailoring the detector to specific applications. For instance, while NbN and NbTiN are used in applications such as quantum key distribution (QKD) due to their high count rate, low jitter and high efficiency~\cite{Grunenfelder2023}, WSi-based SNSPDs are employed where extremely low dark counts and high IDE at longer wavelengths are mandatory, as dark matter experiments~\cite{Baudis2025}. Therefore, investigating alternative superconducting materials is crucial for improving current technologies and exploring new applications.

In this context, tungsten-based amorphous superconductors represent a promising platform to be explored, with WSi established as a key material for application beyond telecom wavelength up to the mid infrared regime~\cite{Taylor2023} and WGe-based SNSPDs which recently exhibited IDE up to $29\text{ }\mu$m~\cite{Hampel2026}. Fundamental studies have demonstrated the reason for the high efficiency of W-based detectors in the mid-infrared regime, as they enable efficient conversion of the incident photons energy into the electronic system~\cite{Ercolano2025, Chen2026}. {Moreover, elemental W itself has recently attracted attention as SNSPD platforms}~\cite{Chen2026}. {Fundamental studies of disordered elemental W have shown that amorphization leads to more robust superconductivity}~\cite{Hofer2019,Hofer2023,Bagwe2024}, {motivating the exploration of nitrogen-induced disorder as a tuning strategy beyond elemental metals. Building on this, elemental-W-based SNSPDs were recently reported, offering tunable $T_{\rm c}$, high normal-state resistivity ($\rho$), and compatibility with established nanofabrication processes}~\cite{Ma2025}. 

At the same time Re-based SNSPDs are also gaining attention in this field. For instance, noncentrosymmetric Nb$_{0.18}$Re$_{0.82}$ films were successfully implemented in both fast nano- ~\cite{Cirillo2020} and microwire~\cite{Ejrnaes2022,Ercolano2023} single-photon detectors, with detection efficiencies up to to 2~$\mu$m~\cite{Carla2024} working at easily accessible operating temperatures of Gifford-McMahon (GM) closed-cycle cryostats. Similarly to the case of tungsten, the addition of nitrogen in a reactive deposition process, produces Nb$_{0.18}$Re$_{0.82}$ nitride films, which resulted in SNSPDs with improved performances both in terms of IDE and response times even at larger temperatures~\cite{Francesco2025}.

Building on this, we investigate tungsten-rhenium superconducting alloys as a new SNSPDs platform. W$_{x}$Re$_{1-x}$ alloys, first studied in bulk form in the 1960s and 1970s~\cite{Federer1965,Easton1974}, exhibit multiple stable and metastable phases depending on composition and deposition conditions. W–Re alloys offer wide tunability of both normal- and superconducting-state properties through composition and growth conditions~\cite{Federer1965,Easton1974}. {In particular, W$_{0.75}$Re$_{0.25}$ (hereafter WRe) emerges as a promising platform capable of bridging the gap between crystalline and amorphous SNSPDs, combining elevated operating temperatures and fast relaxation dynamics with the typical structural advantages and long-wavelength efficiency potential of amorphous materials.} 

{While elemental W, especially in the nitrogen-doped form,}~\cite{Hofer2019,Hofer2023,Bagwe2024,Ma2025,Chen2026} {is already a compelling SNSPD platform, sputtered WRe films emerge as an even more potentially interesting material, due to their higher $\rho$ and $T_{\rm c}$}~\cite{Fra2025,Abhishek_SUST_2026,Abhishek_Nanoscale_2026}. Moreover, as in W, nitrogen incorporation during deposition can be used to further tune the material properties, increasing both $\rho$ and $T_{\rm c}$, and promoting amorphous growth under appropriate deposition conditions. The resulting amorphous films behave as dirty superconductors with moderate $T_{\rm c}$ $\approx$ 6 K and small superconducting coherence length, $\xi$ $\approx4$~nm ~\cite{Fra2025}. In addition, quasiparticle relaxation processes, examined through fundamental magnetoconductivity studies, suggest fast characteristic times and a potentially good detection capability at long wavelengths~\cite{Abhishek_SUST_2026,Abhishek_Nanoscale_2026}.

\begin{figure}[!h] 
    \centering
    \subfloat
    {\includegraphics[width=7cm]{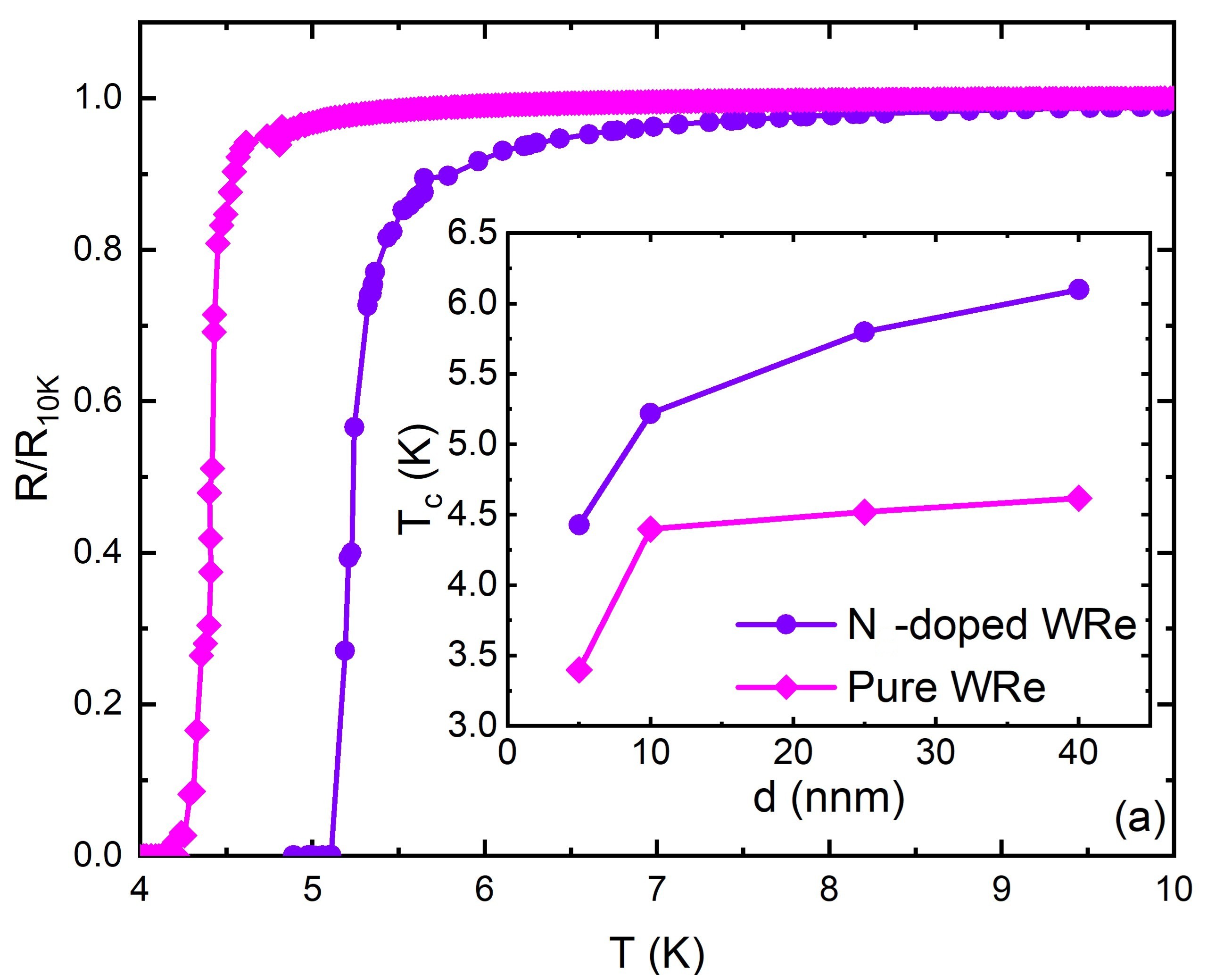}}\quad
    {\includegraphics[width=7cm]{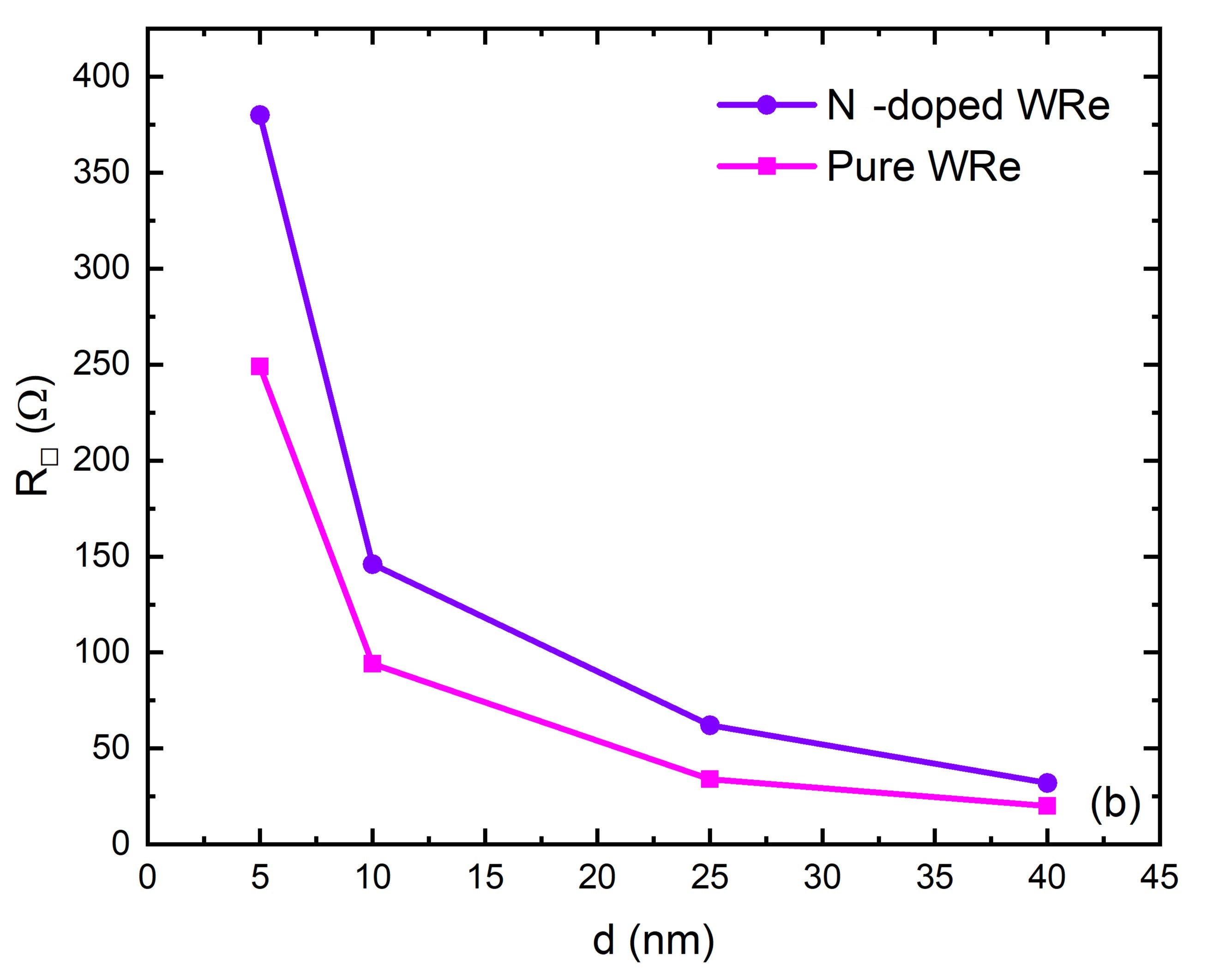}} 
    \caption{(a) Normalized $R(T)$ for N-doped (violet) and pure (magenta) 10-nm WRe films with $T_\mathrm{c}(d)$ as an inset; (b) $R_\mathrm{\square}(d)$ of the two samples (same color scheme) at $10$~K.}  
    \label{RT}
\end{figure}

\begin{figure*}[] 
    \centering
    {\includegraphics[width=15cm]{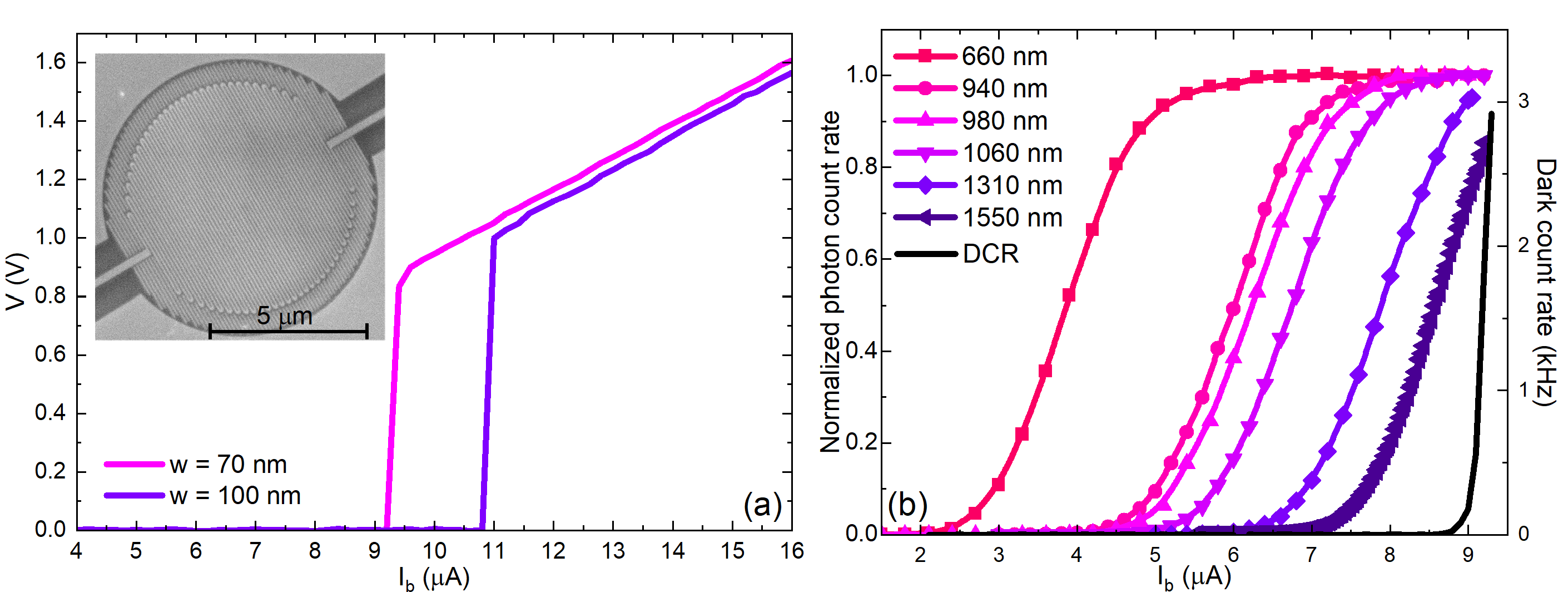}}
    \caption{(a) $V(I_b)$ characteristic of two N-doped WRe SNSPDs (linewidth of $w=70$ and $100$~nm) with inset showing the SEM picture of the $w=70$~nm detector; (b) normalized photon count rate of the same $70$-nm linewidth N-doped WRe SNSPD as function of $I_b$ at different laser wavelength (dots) and their sigmoidal fits (lines).}  
    \label{IV_IDE}
\end{figure*}

Here we demonstrate SNSPDs fabricated from thin amorphous WRe films doped with nitrogen. By controlling sputtering parameters, we realize 10-nm-thick layers with $T_{\rm c}$ suitable for standard cryogen-free operation. Nanowire meanders 70~nm wide with a $9$~µm active-area diameter achieve internal detection efficiency {exceeding 95\% and 85.3\% at 1310 and 1550~nm, respectively,} together with promising timing performance at 2.5~K, establishing this material as a competitive and versatile addition to the SNSPD materials landscape.

\section{Methods}

{The films were deposited via ultra-high vacuum magnetron sputtering (base pressure in
the low $10^{-8}$~mbar range) from a W$_{0.75}$Re$_{0.25}$ target (Testbourne, 99.99\%
purity) at 150~W power both in Ar and Ar/N$_2$ mixture atmosphere. The Ar pressure was fixed to $P_\mathrm{Ar}=3.0$~$\mu$bar, while, in the case of Ar/N$_2$ mixture, the N$_2$ gas flow was set as the 7.5\% of the total incoming gas flow by two separate mass flow controllers. Films' thickness was monitored via a quartz crystal microbalance calibrated with a Bruker DektakXT profiler.} {At this specific percentage, N$_2$ gas was demonstrated}~\cite{Fra2025,Abhishek_Nanoscale_2026} {to get incorporated in the WRe as intestitial atoms, increasing the disorder and leading to an amorphous phase (see morphological characterization in Refs}~\cite{Fra2025,Abhishek_SUST_2026,Abhishek_Nanoscale_2026}). {As shown in Ref.}~\cite{Fra2025}{, this deposition procedure leads to the following film composition: (68.9 ± 0.3)\% of tungsten,  (22.1 ± 0.4)\% of rhenium, and (9.0 ± 0.7)\% of nitrogen.}

Resistance ($R$) measurements were performed in a cryogen-free high field (7~T) measurement system by CRYOGENIC, Ltd, using a $10$~µA excitation current via a power supply Keithley 6121 and a nanovoltmeter Keithley 2182 operating in Delta mode. {The superconducting resistive transitions of the 10-nm samples are reported in Fig.}~\hyperref[RT]{\ref*{RT}(a)}, {with the} $T_\mathrm{c}(d)$ {trend displayed as an inset. Specifically, the $10$-nm N-doped WRe films exhibit} $T_\mathrm{c}=5.22$~K, almost one degree {larger compared to the $4.42$~K found for the films deposited in pure Ar environment. From $R(T)$ measurements, resistivities of} $\rho^{10K}=93.6$ and $146$~µ$\Omega$cm {were extracted at $T=10$~K via the van der Pauw method}~\cite{Pauw1958,Koon1992} {for the $10$-nm pure and N-doped WRe films, respectively. The sheet resistance} $R_\mathrm{\square}$ {of the two film series is also reported as a function of $d$ in Fig.}~\hyperref[RT]{\ref*{RT}(b)}. These results prove the increase in $T_\mathrm{c}$ and $\rho$ caused by the nitrogen doping, {which are beneficial for SNSPD applications. Furthermore, the Nitrogen-doping also shortens phase relaxation time and increases electron-phonon to electron-electron relaxation time ratio}~\cite{Abhishek_SUST_2026}. {Among the deposited films, the 10-nm ones were selected for the current study due to the favorable $T_\mathrm{c}$, which allowed working at $T=2.5$~K while keeping the ratio $T/T_\mathrm{c}>0.5$.}

{As reported in previous works}~\cite{Fra2025,Abhishek_SUST_2026,Abhishek_Nanoscale_2026}, {we attribute the observed behavior to a thin, intrinsically amorphous sub-layer already present in the undoped WRe films, analogous to the disorder-driven superconducting layer identified in elemental W by Bagwe et al}.~\cite{Bagwe2024}. {Interstitial N atoms incorporate preferentially within this disordered region, extending it through a progressively larger fraction of the film thickness, further enhancing both $T_\mathrm{c}$ and $\rho$. Magnetoconductance analyses}~\cite{Abhishek_SUST_2026,Abhishek_Nanoscale_2026} {further indicate enhanced electron-phonon coupling within this increasingly amorphous phase, consistent with the observed $T_\mathrm{c}$ trend.}


Superconducting nanowire single-photon detectors were patterned via a $100$~kV electron beam lithography system (Raith EBPG-5200) on a $100$-nm-thick AR-P~6200 resist. Subsequently, structures were etched using reactive ion etching with a gas mixture of {SF$_6$}/{O$_2$} ($13.5/3.5$~~sccm flow rates). After the fabrication, the remaining resist was removed using wet cleaning (anisole) and then SNSPDs were covered by a $12$~nm layer of SiN deposited via plasma-enhanced chemical vapor deposition, to prevent the film degradation (e.g., oxidation). A scanning electron micrograph of a representative detector is shown in the inset of Fig.~\hyperref[IV_IDE]{\ref*{IV_IDE}(a)}.

The chips were mounted in a GM closed-cycle cryostat operating at a base temperature of $T=2.5$~K, and characterized under flood illumination. Nanowires with $70$~nm linewidth, $140$~nm pitch, and $4.5$~$\mu$m radius yielded the highest IDE and the shortest response time. This layout was therefore adopted as the reference design throughout the manuscript.

\section{Results}

The critical current ($I_{\mathrm c}$) was determined by voltage (V) vs bias current ($I_{\mathrm b}$) measurements, which led to $I_{\mathrm c}=9.4$~µA, corresponding to a critical current density $J_{\mathrm c}=1.34\times10^{10}$~A/m$^2$ at $T=2.5$~K. The same $J_\mathrm{c}$ values were measured on devices with $100$~nm linewidth, confirming the robustness of the film and fabrication process. The measurements on both devices are displayed in Fig~\hyperref[IV_IDE]{\ref*{IV_IDE}(a)}. The depairing current $J_{\mathrm{dp}}(0)=1.57\times10^{11}$~A/m$^2$ at $T=0$~K was evaluated using the formula $J_\mathrm{dp}(0) = \frac{8\pi^2\sqrt{2\pi}}{21\,\zeta(3)\,e}\left[\frac{(k_B\,T_c)^3}{\hbar\,v_F\,\rho\,(\rho\,l)}\right]^{1/2}$ of Ref.~\cite{Romijn1982}, where $\zeta(3)$ is the Apéry constant, $v_F$ the Fermi velocity and $l$ the mean free path. The product $v_F\,l$ was evaluated from the diffusion constant formula $D=v_F\,l/3$~\cite{Caputo2017}, where $D$ was obtained from Ref.~\cite{Abhishek_Nanoscale_2026}. The value of the depairing current at the operation temperature of 2.5 K was then calculated via the Kupriyanov and Lukichev (KL ) universal theory for the ratio $\left[\frac{J_{\mathrm{dp}}(T)}{J_{\mathrm{dp}}(0)}\right]^{2/3}$~\cite{Kupriyanov1980}. Specifically, the numerical solution of the KL theory led to $J_{\mathrm{dp}}(T=2.5\,\text{K})=4.23\times10^{10}$~A/m$^2$. {It is worth noticing that ${J_{\mathrm{c}}}/{J_{\mathrm{dp}}}=0.32$ at $T=2.5\text{ K}$, indicating a non-negligible suppression of the critical current relative to the depairing limit, likely related to  constrictions in the nanowire}~\cite{Frasca2019}.

Photon count rates and dark count rates of the $70$-nm linewidth SNSPD as a function of bias current $I_b$ at $T=2.5$~K are shown in Fig.~\hyperref[IV_IDE]{\ref*{IV_IDE}(b)} for wavelengths ($\lambda$) of $660\text{, }940\text{, }980\text{, }1060\text{, }1310\text{, and }1550$~nm under flood illumination, together with the corresponding sigmoid fits. {The data were normalized to the upper asymptote of the corresponding sigmoidal fit. For $\lambda$ up to $1310$ nm, the count rate reach the $(95.2\pm0.7)$\% of the saturation values of the corresponding sigmoid fitting functions, resulting in saturated IDE. In contrast, the $1550$~nm curve reaches $(85.3\pm0.6)$\% of the sigmoid asymptotic saturation value. The uncertainties on normalized data were propagated from the standard error of the fitted asymptote.}

\begin{figure}[h] 
    \centering
    {\includegraphics[width=8.2cm]{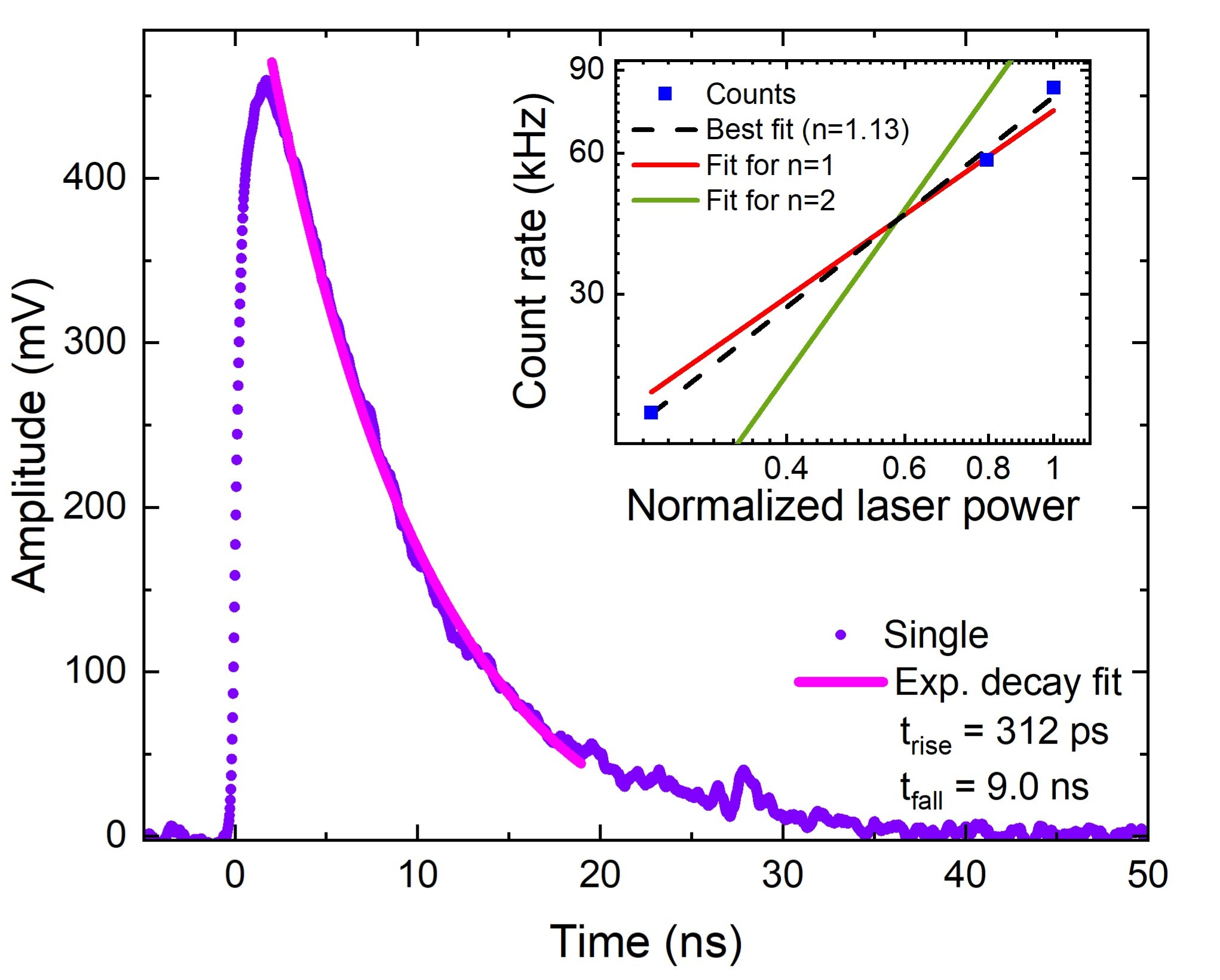}} 
    \caption{Single SNSPD pulse data points (violet dots) of the $70$-nm linewidth device, with exponential fit of the decay (magenta line); characteristic times are also reported. In the inset, a log-log plot of the SNSPD count rate versus normalized laser power. The experimental data (blue squares) are fitted with a linear regression (dashed black line). Fixed-slope reference lines for $n = 1$ and $n = 2$ (red and green solid lines, respectively) are also included.}  
    \label{pulse}
\end{figure}

The temporal response was characterized at $I_{\mathrm b}\approx0.95\,I_{\mathrm c}$ {via a $1060$~nm picosecond ($5.2$~ps) pulsed laser with spectral width of $0.57$~nm.} The output signal, amplified at room temperature, was recorded with a 4~GHz bandwidth oscilloscope. A representative pulse waveform is shown in Fig.~\ref{pulse}, together with an exponential fit to the decay. Defining the rise time between 20\% and 80\% of the pulse amplitude yields $t_\mathrm{rise} = 312$~ps, while the fit provides a fall time $t_\mathrm{fall} = 9.0$~ns. These values already indicate fast detector dynamics for this material. The distortions visible along the trailing edge of the pulse are ascribed to signal reflections in the readout electronics. 

\begin{figure}[!h] 
    \centering
    {\includegraphics[width=8.2cm]{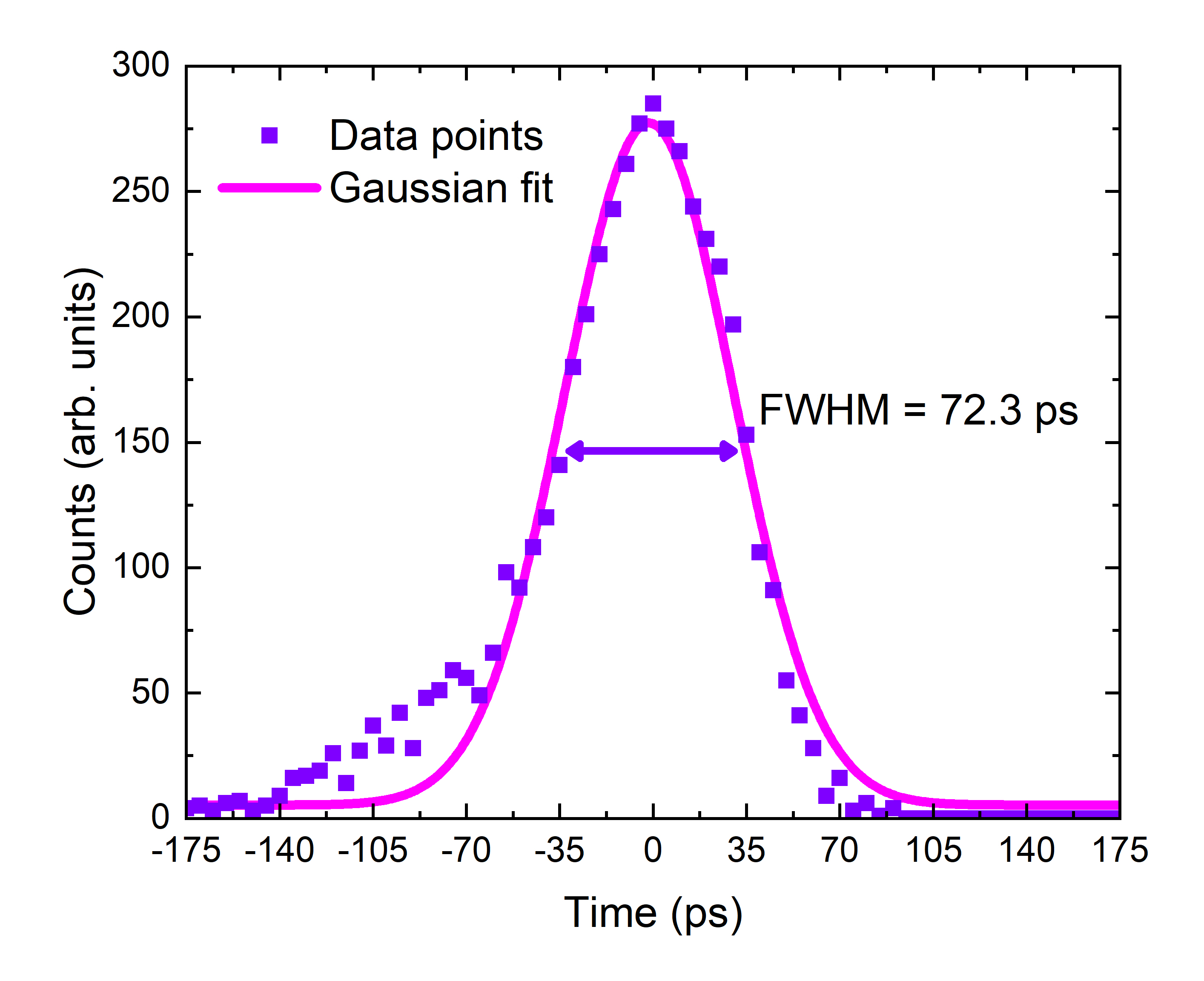}} 
    \caption{$70$-nm linewidth SNSPD timing jitter data points (violet dots), with Gaussian fit (magenta line) and resulting FWHM.}  
    \label{jitter}
\end{figure}

{To verify the single-photon operating regime of the SNSPD, the detector count rate was plotted as a function of the normalized laser power {(absolute power of 1.9~mW at 50~Hz repetition rate)} at $\lambda = 1060\text{ nm}$ on a double-logarithmic scale in the inset of Fig.}~\ref{pulse}. {The attenuation values used in the measurements were 24, 25, and 30~dB, corresponding to approximately $8.1\times10^5$, $6.4\times10^5$, and $2.0\times10^5$ photons per pulse, respectively, on the detector plane.} {The fit to the data}~\cite{Goltsman2001} {yields a slope of $n = 1.13 \pm 0.06$. Reference lines corresponding to slopes of $n = 1$ and $n = 2$. The near-unity exponent, combined with the clear discrepancy between the data and the $n=2$ trendline, confirms that the detection mechanism is dominated by single-photon absorption.}

The timing jitter of the representative device, measured with the same setup, is reported in Fig.~\ref{jitter}. {In our start-stop configuration, the SNSPD pulse defines the START signal and the periodic laser synchronization pulse the STOP, with the zero of the time axis arbitrarily set by the total cable and electronic delay. The distribution exhibits a secondary feature at shorter times, which we attribute primarily to a weak optical echo arising from reflection at the chip surface, consistent with the ~1 cm fiber-to-chip distance. The slight asymmetry can also be enhanced by localized resistivity non-uniformity along the nanowire or photon-detection in the bends}~\cite{Zadeh2020,Sidorova2017}. The full-width half-maximum (FWHM) of the Gaussian fit of the distribution yield to a timing jitter of $72.3$~ps at $T=2.5$~K. This value is already encouraging for an early implementation of the platform and leaves clear room for further reduction through film optimization, refined nanofabrication, and the adoption of cryogenic amplification~\cite{Holzman2019,Zadeh2017}. In fact, patterning-related limitations are observed in a fraction of the devices. Those are consistent with the early stage of process development for this material and highlight significant opportunities for further performance improvements. In this respect, further exploration of the sputtering parameter space—such as the sputtering power and nitrogen partial pressure—is expected to enable finer control over the superconducting properties of the films. Similar optimization strategies have proven effective in related material systems, for instance NbReN, where tuning the plasma I–V characteristics led to marked performance enhancements~\cite{Francesco2025}. Nevertheless, from this study nitrogen-doped WRe thin films emerge as a compelling platform for high-speed single-photon detection. The devices presented here display rapid electrical response, with short rise and decay times, together with an already promising timing jitter.

\section{Discussion}

To benchmark the performance of our N-doped WRe SNSPDs, the characteristics of other material platforms are reported in Table~\ref{tab:comparison}, noting that, since such values were obtained under heterogeneous device geometries and readout conditions, the comparison should be regarded as qualitative. Additionally, due to their dependence on $T$, it is worth clarifying that the relaxation times ratios in Table~\ref{tab:comparison} {were evaluated in proximity of the $T_\mathrm{c}$ of the corresponding materials.} Although mature crystalline nitrides (NbN and NbTiN) benefit from decades of target and geometry optimization, our first-generation devices already challenge new crystalline materials (e.g., NbRe and NbReN) and conventional amorphous systems like WSi, MoSi, or WGe in some crucial aspects. Firstly, such amorphous materials require complex sub-Kelvin cooling ($120$~mK to $1$~K) to reach peak performance, while our devices achieve competitive results already at $T=2.5$~K, accessible via standard closed-cycle cryostats. Additionally, under room-temperature amplification, N-doped WRe exhibits lower timing jitter and reset time compared to some well-established amorphous alternatives like WSi.

Since N-doped WRe explores the same nitrogen-doping strategy investigated in the nitrogen-doped amorphous W platform~\cite{Ma2025,Chen2026}, a direct comparison with this system is particularly instructive. In spite of the not-fully-saturated 1550~nm curve, our devices reach a jitter of $72.3$~ps and a fall time of 9.0~ns under room-temperature amplification, less than half the 175~ps jitter and 22~ns fall time reported for N-doped W, despite the latter benefiting from cryogenic amplification, a configuration intrinsically favorable to lower jitter. It is also worth to underline that this advantage is obtained at an operating temperature of $T=2.5$~K, rather than at $1.0$~K as required by N-doped W. We note that our reported IDE at $1550$~nm (85.3\%) is a non-fully-saturated, fit-extrapolated value, whereas N-doped W reaches saturated unity efficiency at this wavelength. Narrowing this gap is a natural direction for the optimization of the present first-generation devices.

\setlength{\tabcolsep}{2pt}
\begin{table*}
\vspace{3mm}
\begin{centering}
\caption{Performance comparison of the present N-doped WRe SNSPDs with enstablished crystalline (C) and amorphous (A) competitor platforms.}
\label{tab:comparison}
\begin{tabular}{c  c  c  c  c  c  c  c  c}
\hline
\hline
Material & d (nm) & Jitter (ps) & $t_\mathrm{fall}$ (ns) & {IDE at 1550 nm} & Phase & $T$ (K) & $\tau_\mathrm{e-ph}/\tau_\mathrm{e-e}$ & $L_\mathrm{k,\Box}\,(\mathrm{pH}/\Box)$ \tabularnewline
\hline 
N-doped WRe$^*$\cite{Abhishek_Nanoscale_2026} & $10$ & 72.3$^\ddagger$ & 9.0 & 85.3\% & A & 2.5 & 1.03-1.51 & 67.3 \tabularnewline

NbTiN~\cite{Zadeh2017,Miki2009}& $3.5\text{ - }8.4$ & 14.8-48.8$^\dagger$ & 20.3 & 100\% & C & 2.5 & - & 27 - 35 \tabularnewline

NbN~\cite{Hu2020,Incalza2026,Sidorova2020,Miki2009}& $3.5\text{ - }6$ & 66$^\dagger$ & 1.2-42 & 100\% & C & 2.1-4.2 & 0.86 & 33 - 50 \tabularnewline

WSi~\cite{Marsili2013,Verma2015,Zhang2016}& $4.5\text{ - }5$ & 150$^\ddagger$ & 40 & 100\% & A & 0.120-2 & 1.4-3.8 & 250 - 350 \tabularnewline

MoSi~\cite{Caloz2018,Verma2015,Ercolano2025}& $5\text{ - }7$ & 26.1-44.2$^\dagger$ & 35 & 87\% & A & 0.8 & 1 & 190 \tabularnewline

NbRe~\cite{Cirillo2020}& $8$ & 33.1$^\dagger$ & 8-19 & 89\% & C & 2.3-3.3 & - & 139 \tabularnewline

NbReN~\cite{Francesco2025}& $10$ & 28.4$^\dagger$ & 8.0 & 95\% & C & 3.5 & - & 83.8 \tabularnewline

WGe~\cite{Yang2025,Ercolano2025}& $5\text{ - }10$ & 127-234$^\ddagger$ & - & 100\% & A & 0.4 & 2.6 & - \tabularnewline

N-doped W~\cite{Chen2026}& $8$ & 175$^\dagger$ & 22 & 100\% & A & 1.0 & - & - \tabularnewline

MoGe~\cite{Verma2014_022602,Ercolano2025}& $5\text{ - }7.5$ & 72$^\ddagger$ & 9.0 & 100\% & A & 1.0 & 1.2 & 61 \tabularnewline

MoRe~\cite{Nevzorov2025}& $4$ & 132$^\ddagger$ & 4.1 & 73.5\% & A & 2.5 & - & - \tabularnewline

\hline  
\hline

\end{tabular}
\par\end{centering}
{\raggedright
$\qquad^*$ present work\\
$\qquad^\dagger$ cryo-amplifier\\
$\qquad^\ddagger$ room temperature amplifier\\}
\end{table*}


{Moreover, based on the experimental $t_\mathrm{fall}$ and the readout circuit impedance, the sheet kinetic inductance was evaluated as $L_{\mathrm{k},\Box} = 67.3\text{ pH}/\Box$. Interestingly, in spite the amorphous nature of the film, this remarkably low value is closer to that of crystalline NbN or NbTiN}~\cite{Miki2009} {than to the typical $250-350\text{ pH}/\Box$ amorphous WSi}~\cite{Marsili2013,Verma2015}. {Therefore, N-doped WRe demonstrates accessible operating temperatures and fast relaxation dynamics typical of crystalline nitrides, while preserving the structural advantages of an amorphous material, creating a bridge between the two different classes of materials.}

The potentialities of this material are also supported by previous fundamental analysis {on both pure and Nitrogen-doped WRe films} deposited in similar conditions~\cite{Fra2025,Abhishek_SUST_2026,Abhishek_Nanoscale_2026}. In particular, non-equilibrium processes were performed through magnetoconductance experiments, from which fast thermalization times occurring both by electron-electron ($\tau_\text{e$-$e}$) and electron-phonon ($\tau_\text{e$-$ph}$) interactions were estimated~\cite{Abhishek_SUST_2026,Abhishek_Nanoscale_2026}. Both the values of $\tau_\text{e$-$e}$ and $\tau_\text{e$-$ph}$ are, indeed, smaller compared to other W-based amorphous films (such as WSi), as well to Mo-based films (MoGe and MoSi), which are typically characterized by fast dynamics~\cite{Abhishek_SUST_2026,Abhishek_Nanoscale_2026,Ercolano2025}. Moreover, the same studies suggest an efficient energy transfer from the photon to the electronic system, as measured by the ratio $\tau_\text{e$-$ph}/\tau_\text{e$-$e}$, as extensively explained in Ref.~\cite{Ercolano2025}. Indeed, for a 5-nm-thick nitrogen-doped WRe bridge it is $\tau_\text{e-ph}/\tau_\text{e-e}$ $\approx$1.5~\cite{Abhishek_Nanoscale_2026}. This value can be compared to the results obtained using the same approach for other amorphous films of similar thickness in Table~\ref{tab:comparison}. In fact, it results that $\tau_\text{e-ph}/\tau_\text{e-e}$ is intermediate between the large ratios reported for W-based films [for example $(\tau_\text{e-ph}/\tau_\text{e-e})^{WSi}=2.5$] and Mo-based systems [for instance $(\tau_\text{e-ph}/\tau_\text{e-e})^{MoSi}=1.0$]~\cite{Ercolano2025}. Crucially, these results suggest that further scaling down the film thickness could significantly boost sensitivity, noting that typical amorphous counterparts operate in a much thinner regime (Table~\ref{tab:comparison}).

\section{Conclusions}

In conclusion, this work extends the range of materials used for the established SNSPD technology to nitrogen-doped WRe. Introducing nitrogen during deposition is an effective strategy to realize amorphous films and enhancing at the same time both the resistivity and $T_\mathrm{c}$. The resulting devices show competitive performance balancing efficiency, speed and high operating temperature, such as $85.3\%$ IDE at 1550 nm and sub-100 ps timing jitter at 2.5 K. This valuable preliminary results suggest that the metrics reported in this first implementation can be further improved for instance reducing the films thickness, optimizing their growth, also in terms of nitrogen doping, and, finally, refining the nanofabrication process.





\data{The data that support the findings of this study are available from the corresponding author upon reasonable request.}


\bibliography{bibliography}

\end{document}